# Experimental transmission of quantum digital signatures over 90-km of installed optical fiber using a differential phase shift quantum key distribution system

ROBERT J. COLLINS,[1,*] RYAN AMIRI,[1] MIKIO FUJIWARA,[2] TOSHIMORI HONJO,[3] KAORU SHIMIZU,[3] KIYOSHI TAMAKI,[3] MASAHIRO TAKEOKA,[2] ERIKA ANDERSSON,[1] GERALD S. BULLER,[1] MASAHIDE SASAKI[2]

[1]*Institute of Photonics & Quantum Sciences, and the Scottish Universities Physics Alliance, David Brewster Building, Gait 2, Heriot-Watt University, Edinburgh EH14 4AS, United Kingdom;*
[2]*Quantum ICT Laboratory, National Institute of Information and Communications Technology (NICT), 4 2 1 Nukui Kitamachi, Koganei, Tokyo 184-8795, Japan*
[3]*NTT Basic Research Laboratories, NTT Corporation, 3-11 Morinosato Wakamiya, Atsugi, Kanagawa 180-8585, Japan*
**Corresponding author: r.j.collins@hw.ac.uk*

**Quantum digital signatures apply quantum mechanics to the problem of guaranteeing message integrity and non-repudiation with information-theoretical security, which are complementary to the confidentiality realized by quantum key distribution. Previous experimental demonstrations have been limited to transmission distances of less than 5 km of optical fiber in a laboratory setting. Here we report the first demonstration of quantum digital signatures over installed optical fiber as well as the longest transmission link reported to date. This demonstration used a 90-km long differential phase shift quantum key distribution system to achieve approximately one signed bit per second – an increase in the signature generation rate of several orders of magnitude over previous optical fiber demonstrations.**

For many years, it was generally trusted that a simple handwritten signature was hard to forge and, therefore, could be taken as certification that the signatory was willing to be considered a source of the message and agreed with the contents. In addition, there was an acceptance that a signed message would be validated by a third party, which allowed complex transaction systems to operate with reasonable ease.

The field of digital signatures seeks to restore some of the practical security aspects of handwritten signatures lost in the transition to digital communications. Digital signatures must guarantee no forging (that a message is signed by the legitimate sender and has not been modified), and non-repudiation (Alice cannot repudiate her message, that is, successfully deny that she sent it). Currently used digital signature schemes typically rely on "one-way" functions, which are computationally easy to evaluate in one direction but computationally difficult to invert without additional information. However, there is no existing proof of the long-term security of these signature schemes and they are vulnerable to algorithmic breakthroughs, emerging quantum data processing technologies and even significant large-scale investment in conventional computational technologies, – thus they only offer "computational security" against an attacker with currently accepted standards of reasonable computational resources.

Naturally, there is strong motivation to develop digital signature schemes that offer "information-theoretical security", meaning that they are secure against all attackers, even those with unlimited computational resources Wegman-Carter message authentication offers information-theoretical security. However, to guarantee non-repudiation as well, one possibility is to use quantum digital signatures. QDS schemes operate using technology similar to that of QKD but employ protocols that do not require distillation of a fully secret key. Indeed, it is possible that both QKD and QDS schemes can work in parallel along the same optical fibers. This paper demonstrates the first implementation of QDS in an installed optical fiber infrastructure, as well as the longest QDS link reported so far.

The first experimental demonstration of QDS operated over a limited transmission distance of 5 m, using an all-optical multiport to guarantee transferability of the phase-encoded coherent states that comprised the signature [1]. The all-optical multiport was effectively two interwoven optical fiber Mach-Zehnder interferometers with the inherent stabilization issues such a design entails and exhibited relatively high optical loss of 7.5 dB. Furthermore, this first experimental demonstration required a quantum memory at each receiver to store the optical coherent states until the classical description was transmitted. This transmission could occur a significant time after the quantum transmission, therefore rendering this protocol impractical.

The requirement for quantum memory was removed in a subsequent revision to the protocol which employed unambiguous state elimination (USE) measurement of the optical coherent states at the receiver [2]. This realization was still limited to transmission ranges of 5 m by the multiport. A further revision to the protocol removed the requirement for the multiport [3], permitting operation at transmission distances of up to 2 km [4]. The primary limit on the transmission distance of this system was the comparatively high loss of the optical fiber, 2.2 dBkm$^{-1}$ at the wavelength of 850 nm used for these experiments. This wavelength was chosen to offer compatibility with low-noise semiconductor silicon single-photon avalanche diode (Si-SPAD) single-photon detectors [5] at the expense of higher transmission channel loss. Recent developments in semiconductor [6] and superconducting [7] single-photon detection technologies offer the prospect of low-noise operation at wavelengths in the traditional telecommunications wavelength bands around 1550 nm.

Digital signature protocols have two stages, a distribution stage and a messaging stage. The distribution stage establishes signatures to enable signed messages to be sent and received in the messaging stage, which is entirely classical and could occur at any future date. Fig 1 shows a basic schematic of the operation of the revised protocol used in the work reported in this letter. Distinct from many quantum communications protocols, Bob and Charlie here were senders of photons while Alice was the receiver. Here, we will describe how to sign a 1-bit message, $m$; longer messages may be sent by suitably iterating this procedure.

In the distribution stage, Bob and Charlie independently choose two random sequences of bits, one sequence for each possible message, 0 or 1. In the simple example that follows, we focus on Bob sending a single bit string to Alice for the future message 0. We do this with the understanding that he also does this with a separate bit string for future message 1, and that Charlie also does the same for both single bit messages 0 and 1. To communicate his bit strings to Alice, Bob carries out a partial QKD [8] procedure, without error correction and privacy amplification. Bob subsequently discards any bits for which Alice received no measurement outcome. Bob sends states until Alice has received a pre-specified number of measurement outcomes. Following this, Alice's bit string will be partially correlated with Bob's, but since we do not perform the classical post-processing of QKD, the correlation and secrecy of their bit strings will not be perfect, and an eavesdropper could have partial information.

Alice and Bob then agree on and sacrifice a small number, $k$, of the bits in their keys to estimate the error rate between their strings, leaving a remaining key of length $L$. For each possible future message, Bob and Charlie randomly and independently choose $L/2$ of the bits in their strings to forward to the other recipient. Bob and Charlie keep secret from Alice the bits that are forwarded and the bits that are retained. This last step is achieved using a standard BB84 QKD link (i.e. including error correction and privacy amplification) between Bob and Charlie to provide a secure communications channel.

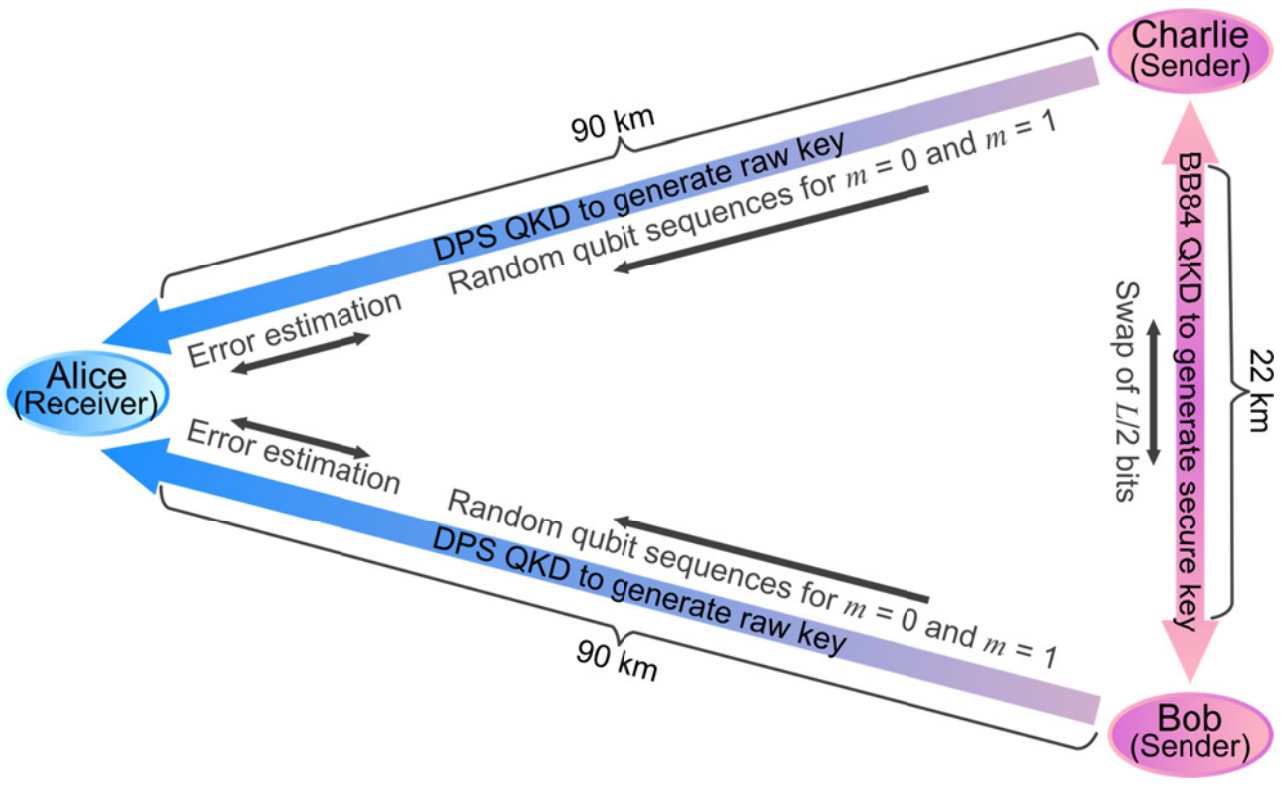

Fig. 1. A schematic representation of the processes carried out to perform the distribution stage of QDS. Bob and Charlie conduct QKD with Alice, without error correction and privacy amplification, to generate partially correlated bit sequences. These are then post-processed over a shared QKD link between Bob and Charlie to generate the quantum digital signatures.

At the end of the distribution stage, for each future message, Alice holds information on the strings sent by both Bob and Charlie, while Bob holds only partial information on the string sent by Charlie, and similarly Charlie holds only partial information on the string sent by Bob.

In the original theoretical outline [3] it was proposed that the communication between Alice and Bob, and between Alice and Charlie, be carried out using phase-basis set QKD with the BB84 protocol conducted to generate a raw key, but without error correction or privacy amplification performed. Here we report the first application of differential phase shift (DPS) QKD [8] to QDS. We employed the 1 GHz clock rate DPS QKD system developed by NTT as part of the Tokyo QKD Network [9,10], as shown in schematic form in Fig 2. In our demonstration, Bob and Charlie alternately performed partial DPS QKD with Alice to generate partly correlated bit strings that can be used for the formation of signatures. Error correction was not required at this stage, as outlined in the protocol described below. DPS QKD has the advantage that there is no requirement for basis set reconciliation and consequently no associated discarding of photon detection events. The system transmitted over a 45-km long fiber in a loopback configuration, as shown in Fig 2, with approximately half of the fiber being in underground ducting and half being overhead lines. The round trip of 90 km of optical fiber exhibited a loss of 31 dB and the Alice receiver optics had an additional loss of 10 dB. The system employed a mean photon number per pulse ($\mu$) at the output of Alice of 0.2 photons per pulse. The superconducting niobium-nitride meander structure nanowire single-photon detectors had a detection efficiency of 30% and a dark count rate, in the absence of signal photons, of 100 counts per second. The combination of these factors meant that the mean detector event rate during operation was approximately 10k counts per second. Bob and Charlie each generated a shared bit string of 2 Mbits with Alice at a quantum bit error rate of 1.08% by transmitting a unique sequence to her. The Bob-Charlie bit exchange was then carried out over a classical channel secured using a separate BB84 QKD system developed by NEC and NICT and operating over a 22-km long optical fiber channel [11].

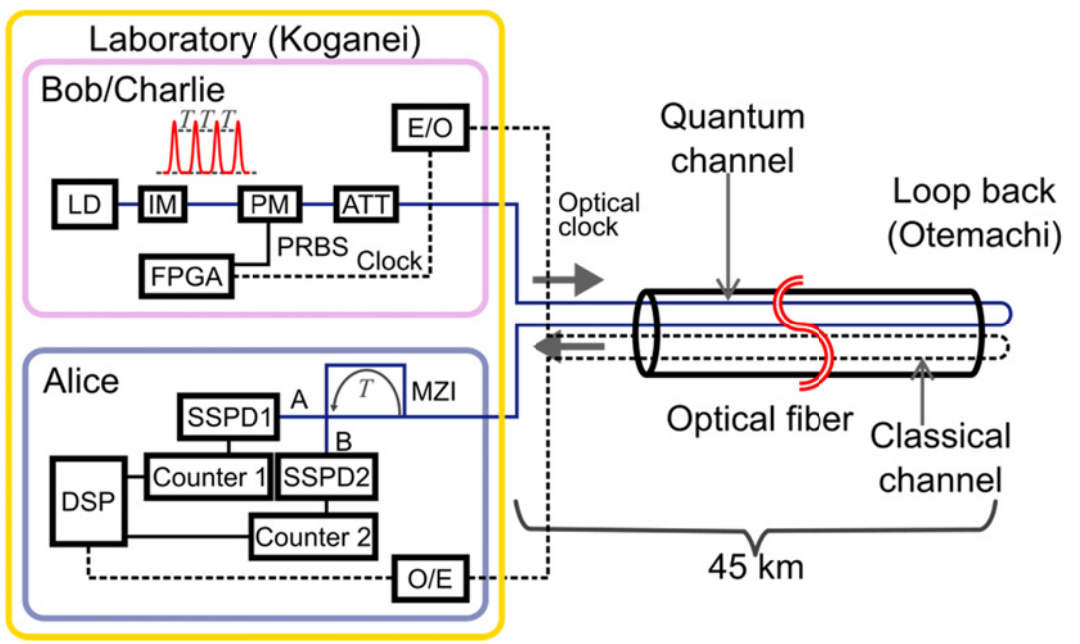

Fig. 2. The DPS system employed to carry out QDS generation. LD is a CW laser diode, IM is an intensity modulator used to generate optical pulses at a clock rate of 1 GHz, PM is a phase modulator used to encode the potential bit value, chosen using a pseudo-random bit sequence (PRBS) from a field programmable gated array (FPGA). The optical attenuator, ATT, selects a desired mean photon level while E/O is an electrical to optical encoder that transmits the clock pulse used to phase lock sender and receiver. At the receiver, Alice, the photons from the sender, Bob or Charlie, were detected by superconducting single-photon detectors (SSPD) and a digital signal processor (DSP) used to compile the data for signature generation. Clock recovery is carried out by the optical to electrical encoder O/E. MZI is a delay of duration $T$, the time between successive pulses introduced by the IM.

In the second stage of the protocol, the messaging stage, Alice chooses a message $m$ and sends it together with her corresponding $2L$ measurement outcomes of the states sent to her by both Bob and Charlie for message $m$. All communication during the messaging stage takes place over pairwise authenticated classical communication channels; quantum communication is needed only in the distribution stage. Naturally, quantum effects must be considered in order to quantify the information obtainable by an eavesdropper in the distribution stage, as this will directly affect the security of the scheme. For the scheme to be robust and secure, we require three security guarantees to hold with high probability. First, we require that if all participants are honest, the protocol does not abort. Second, we require that no one can send a message to Bob or Charlie pretending to be Alice, and get that message accepted. We call this security against forging. Third, we require that Alice cannot send a message that will be accepted by one participant and rejected by the other. We call this security against repudiation/non-transferability. For more exact security definitions, we direct the reader to the recent work of Arrazola *et al.* [12].

Suppose Alice sends the message to Bob. To accept the message, Bob checks Alice's private key against the key he sent for the message $m$. He accepts the message if he finds fewer than $Ls_a$ mismatches, where $s_a$ is an authentication threshold, otherwise he rejects it.

If Bob wishes to forward the message, he sends the message together with Alice's private key to Charlie. Charlie then checks for mismatches in the same way as before but applies a different verification threshold $s_v$, which is greater than $s_a$. The message is only accepted if there are fewer than $Ls_v$ mismatches, otherwise it is rejected. It is important that the threshold for accepting a message directly from Alice is different from the threshold for accepting a forwarded message, otherwise Alice could repudiate with high probability. Signing a message uses up the distributed signatures, which cannot be reused.

In order to provide a security analysis, we use the approach outlined by Diamanti [13] to bound Eve's ability to successfully forge. Namely, for the given experimental set up, we can find Eve's probability of successfully guessing each bit sent by Bob/Charlie to Alice. The results are valid only for a restricted class of attacks by Eve, namely individual attacks and sequential attacks. Individual attacks are attacks in which the eavesdropper acts separately and independently on each qubit sent over the quantum channel. Collective measurements and operations entangling successive qubits are not allowed. Sequential attacks are an extended form of the "intercept and resend" strategy whereby Eve waits for a certain number of consecutive detection events and modifies her strategy accordingly. Though these attacks may seem restrictive, they essentially include all attacks possible with current technology. Further, security in the completely general setting could be proven using the results of Tamaki *et al.* [14], but would require a slightly modified experimental set-up as well as photon number resolving detectors.

Using adapted forms of equations (3.23) and (3.28) of [13], we find that Eve's probability of incorrectly guessing the bit sent by Bob is

$$P_e = 1 - Max\{ 2\mu(1-T) + \left(1-2\mu(1-T)\right)\left(1-e^2-\tfrac{1}{2}(1-6e)^2\right), 2de + \frac{1}{2}(1-2de)\}, \quad \textbf{(1)}$$

where $T$ is the total transmission efficiency of the system, $e$ is the QBER, and $d = \mathrm{Log}_{\mu} T + 1.$, as derived from (3.26) of [13]. For the experimental set-up considered here, we have

$$P_e = 0.262. \quad \textbf{(2)}$$

With this, we can use Hoeffding's inequalities [15], together with the methods in our previous work [3] to find

$$\mathrm{Prob(Honest\ Abort)} \leq 2\mathrm{Exp}[-(s_a - e)^2 L] \quad \textbf{(3)}$$
$$\mathrm{Prob(Repudiation)} \leq 2\mathrm{Exp}[-\left(\frac{s_v - s_a}{2}\right)^2 L] \quad \textbf{(4)}$$
$$\mathrm{Prob(Forge)} \leq \mathrm{Exp}[-(P_e - s_v)^2 L]. \quad \textbf{(5)}$$

We say that the protocol is secure to a level of ε if the above three quantities are all less than ε. To make all probabilities approximately equal (unless we have a reason to favor one over any other) we define $g = P_e - e$ and set $s_a = e + g/4$, $s_v = e + 3g/4$. In our previous demonstrations of QDS, the ε parameter was chosen to be $10^{-4}$. If this is applied to our system, we require $L$ = 2,502 counts at a receiver to sign a half-bit, or approximately two bits could be signed per second. QKD systems typically employ a smaller $\varepsilon$ parameter of around $10^{-10}$, and applying that to our system gives $L$ = 5,992 counts to sign a half-bit (either $m$ = 0 or $m$ = 1) or approximately one signed bit per second.

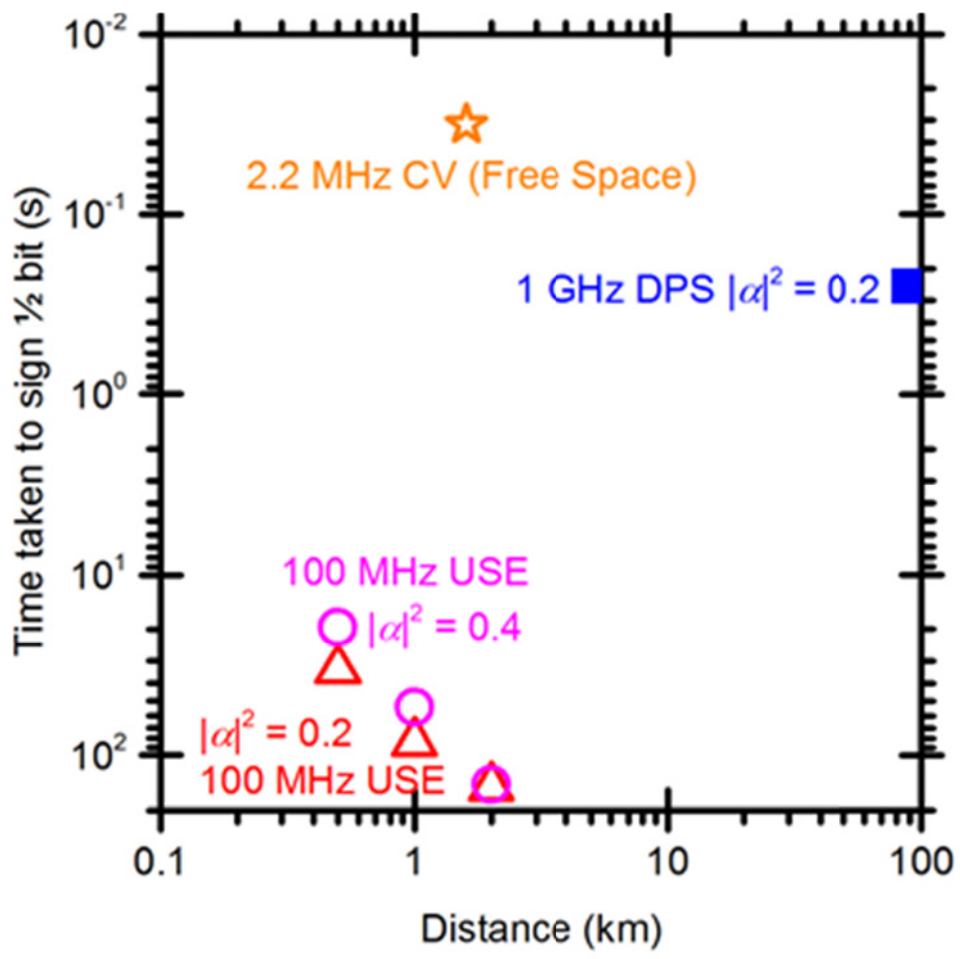

Fig. 3. A comparison between the work presented in this letter (blue filled square), previous optical fiber-based demonstrations employing USE at the same $|\alpha|^2$ of 0.2 (red hollow triangles), previous optical fiber-based demonstrations employing USE at a near optimal $|\alpha|^2$ of 0.4 (magenta hollow circles), and a recent work employing continuous variable QKD to QDS over a free space link (orange hollow star).

We have successfully conducted the first experimental demonstration of quantum digital signatures over 90 km of installed optical fiber. Depending on the desired level of security (as characterized by the parameter $\varepsilon$), the system can sign around one or two bits per second (assuming a value for $\varepsilon$ of $10^{-4}$, and $10^{-10}$, respectively). This compares very favorably, as shown in Fig 3, with our previous demonstrations which, for an $\varepsilon$ of $10^{-4}$, would take either eight years to sign a half-bit [2] in the case of the multiport system or 20 s in the optimum case of the short wavelength system at a distance of 500m [4]. Recent work applying continuous-variable QKD over a 1.6 km free-space link to QDS achieved a signing rate of approximately 33 half bits per second for an $\varepsilon$ of $10^{-4}$ [16]. By experimentally demonstrating that optical fiber QKD systems can be used to provide the additional functionality of signatures, this work widens the field by providing an already established set of experimental resources that can be applied to this complementary security technology, and offers the prospect of commercial systems capable of offering both QKD and QDS, depending on the end application required.

**Funding.** The Daiwa Anglo-Japanese Foundation (10803/11543); The UK Engineering and Physical Sciences Research Council (EPSRC) (EP/G009821/1, EP/K022717/1, and EP/K015338/1); The ImPACT Program of Council for Science, Technology and Innovation (Cabinet Office, Government of Japan).

**Acknowledgment**. The authors gratefully acknowledge K. Yoshino, T. Ochi, and A. Tajima of NEC for supporting the operation of the QKD link between Koganei and Fuchu.